%% file: knauss-hp103-paper.tex
\documentclass[conference]{IEEEtran}
\IEEEoverridecommandlockouts

\usepackage[utf8]{inputenc}
\usepackage{hyperref}
\usepackage{cite}
\usepackage{amsmath,amssymb,amsfonts}
\usepackage{algorithmic}
\usepackage{graphicx}
\usepackage{textcomp}
\usepackage{xcolor}
\usepackage{balance}
\usepackage{fancybox}
\def\BibTeX{{\rm B\kern-.05em{\sc i\kern-.025em b}\kern-.08em
    T\kern-.1667em\lower.7ex\hbox{E}\kern-.125emX}}

\input{macro-nice-tables}

\newenvironment{summarybox}%
{\cornersize{0.15}\setlength{\fboxsep}{0.8em}\begin{center}\vspace{1em}\noindent\begin{Sbox}\begin{minipage}{0.92\columnwidth}}%
{\end{minipage}\end{Sbox}\ovalbox{\TheSbox}\end{center}\vspace{0em}}

\begin{document}

\input{00-frontmatter}
\input{01-intro}
 \input{02-related-work}
\input{03-method}

\input{04-guidelines}
\input{05-results}
\input{06-discussion}

\input{07-conclusion}

\balance

\bibliographystyle{IEEEtran}
\bibliography{knauss-hp103-paper}

\end{document}

%% file: macro-nice-tables.tex
\usepackage{booktabs}
\usepackage{array}
\usepackage{colortbl}
\usepackage{multirow}
\usepackage{longtable}

\newcolumntype{v}[1]{>{\raggedright \hspace {0pt}}p{#1}}
\newcolumntype{G}[1]{>{\columncolor{gray90}}#1}

\definecolor{Gray}{gray}{0.8}
\definecolor{gray25}{gray}{0.25}
\definecolor{gray50}{gray}{0.50}
\definecolor{gray75}{gray}{0.75}
\definecolor{gray90}{gray}{0.9}

\newcommand{\grayrow}{\rowcolor{gray90}}
\newcommand{\graycell}{\cellcolor{gray90}}

%% file: 00-frontmatter.tex
\title{Constructive Master's Thesis Work in Industry: Guidelines for Applying Design Science Research}

\author{\IEEEauthorblockN{Eric Knauss}
\IEEEauthorblockA{\textit{Dept. of Computer Science and Engineering} \\
\textit{Chalmers $\mid$ University of Gothenburg}\\
Gothenburg, Sweden \\
eric.knauss@cse.gu.se}
}

\maketitle

\begin{abstract}
\emph{[Context:]} 
{Software engineering researchers and practitioners rely on empirical evidence from the field. Thus, education of software engineers must include strong and applied education in empirical research methods. For most students, the master's thesis is the last, but also most applied form of this education in their studies. }

\emph{[Problem:]} 
Especially thesis work in collaboration with industry requires that concerns of stakeholders from academia and practice are carefully balanced. 
It is possible, yet difficult to do high-impact empirical work within the timeframe of a typical thesis.
In particular, if this research aims to provide practical value to industry, academic quality can suffer. 
Even though constructive research methods such as Design Science Research (DSR) exist, thesis projects repeatably struggle to apply them. 

\emph{[Principle solution idea:]} 
DSR enables balancing such concerns by providing room both for knowledge questions and design work. 
Yet, only limited experience exists in our field on how to make this research method work within the context of a master's thesis.
To enable running design science master's theses in collaboration with industry, this paper complements existing method descriptions and guidelines. 
It offers experience and pragmatic advice to students, examiners, and supervisors in academia and industry. 

\emph{[Method:]} 
This paper itself is based on DSR. 
Based on 12 design science theses over the last seven years, common pitfalls and good practices are collected from analysing the theses, the student-supervisor interaction, the supervisor-industry interaction, the examiner feedback, and, where available, reviewer comments on publications that are based on such theses.

\emph{[Results:]} 
This paper provides concrete advise for framing research questions, structuring a report, as well as for planning and conducting design science research with practitioners.

\end{abstract}


%% file: 01-intro.tex

\section{Introduction}

The European Qualifications Framework (EQF) structures life long learning in eight distinct levels \cite{SEK+2016}. 
EQF-Level 7 relates to Master's of Science degrees, for which EQF suggests learning outcomes relating to (i) highly specialized knowledge, partly ``at the forefront of knowledge in a field of work or study'', (ii) ``specialized problem-solving skills required in research and/or innovation'', and (iii) the competence to ``manage and transform work or study contexts that are completely unpredictable and require new strategic approaches'' \cite{SEK+2016}.

The master's programs \emph{Software Engineering and Technology} at Chalmers and \emph{Software Engineering and Management} at University of Gothenburg are typical examples of EQF-level 7 programs. 
In their learning outcomes, they explicitly list the skill to \emph{``contribute to research and development in software engineering''} and the ability to \emph{``reflect on the possibilities and limitations of software engineering research, its role in society and the responsibility of the individual for how it is used''}.
These particular learning outcomes are to a large degree to be covered in the master's thesis course, where students are supposed to make a research contribution, most often based on empirical data from companies. 
Both programs are taught at the Department of Computer Science and Engineering and share a small group of examiners for the master's theses of both programs. 
Typically, examiners judge the quality of a master's thesis based on the following high-level criteria:

\begin{enumerate}
\item \textbf{Challenges:} Clarity, relevance to the master's program, and significance.
\item \textbf{Context:} Describe problem and thesis in relation to state of the art and professional literature.
\item \textbf{Methodology:} Quality and fitness of scientific principles of the design and evaluation. 
\item \textbf{Contribution:} Solution completeness and progress beyond state of the art.
\item \textbf{Other:} Quality of writing, work planning and independence, observing ethics and sustainability. 
\end{enumerate}

These criteria have evolved over time, and a dialogue with the examiner is needed for each thesis.
When asked about minimum requirements for passing grades, examiners have said things like \emph{``In a qualitative exploratory study, we expect 10 interviews of 60 minutes, recorded and transcribed, to form the data. 
Different methods should have comparable amounts of data and effort.''} or \emph{``An acceptable thesis should be possible to publish at a conference, for a pass-with-honors, it should be possible to publish it at a journal.''}
This shows the emphasis on the third criteria, the methodology.

These requirements may be seen as in conflict with some master's thesis offerings at local companies, which often relate to practical problems that they would like students to address.
Students are frequently faced with balancing a trade-off between academic values and practical relevance.
More often than not, this trade-off can be very successfully balanced with the help of an experienced supervisor. 
However, there is little guidance to students and they might not be able to initiate such balancing at the right point in time when deciding about a thesis. 
From a pedagogical point of view, the \textbf{constructive alignment} \cite{Biggs2011} of learning goals (asking for research and practical software engineering knowledge) and assessment criteria (mainly focussing on research contribution) is not as clear as desirable in the face of a concrete assignment in industry.

{Many guidelines for empirical research exist} \cite{Creswell2017,Ralph2021}, for experiments \cite{Wohlin2012} and case studies \cite{Yin2014}. 
In particular qualitative exploratory studies based on interviews \cite{Runeson2009} and coding \cite{Saldana2015} have been successfully applied  to explore a new topic or to understand a research phenomenon in a practical setting.
{Yet, there is a demand for less analytical and more constructive master's thesis work as well as for guidelines that complement such more general empirical methods.}

For a certain type of problems, 
the design science research methodology {is an excellent fit} \cite{vaishnavi,wieringa2009,hevner,Engstrom2020}. 
Design science research is a constructive research methodology, which focuses on solving a concrete problem through the design of a tangible design artifact (e.g. a piece of software), while also exploring so-called knowledge-questions. 
In this way, it allows to work constructively on a relevant contribution towards a solution which is relevant to a company partner, and at the same time to collect empirical data about knowledge questions, which often are of high relevance to academia. 

{In this paper,} a master's thesis that constructs a solution to a practically relevant problem {is referred to} as \emph{constructive master's thesis}. 
For such constructive master's theses, design science research is very applicable and valuable.
Yet, students and supervisors find it very difficult to apply.
Through painful experience, 
re-occurring pitfalls and practices to mitigate them{ were discovered}. 
{This paper compiles} these experiences to support students and supervisors in getting most out of their design science thesis work. 

{Based on design science research, experiences were systematically collected}  from {12} master's theses in which the 
author was involved as a supervisor over the last {seven} years
. 
The artifact reported on in this paper are \emph{guidelines to conduct a design science master's thesis in industry} and aims to complement existing method descriptions \cite{vaishnavi,Engstrom2020}, guidelines \cite{hevner}, and practices \cite{wieringa2009}. 
The knowledge questions 
{answered in this paper} are the following:

\begin{description}
\item[RQ1] Which problems are frequently encountered when conducting a master's thesis in industry based on design science research methodology?
\item[RQ2] Which potential solutions exist to mitigate problems with running a design science master's thesis?
\item[RQ3] To what extent can typical problems with running a design science master's thesis be addressed?
\end{description}

{These research questions are answered} based on analysing notes and memories from the interaction between supervisor and student, feedback from examiners and industry partners, and, where applicable, review comments from reviewers of publications that were derived from the theses.
{Likely, these guidelines are valuable beyond the scope of investigation, i.e. the two universities and programs, but clearly the research method relates to this particular context.}

The rest of this paper is organized as follows. 
{The next section describes the research context in terms of background and related work.}
Section \ref{sec:research-method} 
{gives} more details about the research method 
used for this paper. 
Section \ref{sec:guidelines} 
{presents} the artifact: \emph{guidelines to conduct a design science master's thesis in industry}. 
{In particular, the section proposes good practices and shares common pitfalls that should be avoided.}
These guidelines are based on results with respect to 
{the} research questions
{ and constructed}  through 
{the} research method, {for }which 
{results are presented} in Section \ref{sec:results}. 
{Section \ref{sec:discussion} concludes} the paper with a discussion of implications and potential future work.

%% file: 02-related-work.tex
\section{Background and related work}
\label{sec:background}
When reflecting about teaching, Brookfield suggests four different lenses: students eyes, colleagues' perception, theory, and autobiography  \cite{Brookfield1995}.
{T}his section
{ explores} the theoretical foundation
{, while the guidelines and suggestions relate} to students eyes and colleagues' perception
{. In particular, they take} into account four different stakeholder groups: the students, the academic supervisors, the academic examiners, and, in addition to Brookfield's lenses, the industry partners and especially the industry supervisor or champion. 
Through systematically discussing {seven} years of experience with constructive master's theses, 
the autobiographical dimension{ is covered as well}. 
For this purpose, 
{I} rely on self-reflection \cite{Schon1983,Wang2014,CognitiveApprenticeship}.
In particular,
{ my} experience rests on reflections {(often together with other stakeholders)} after each year on which advice was helpful and which practices did yield a good effect. 
Sch{\"o}n argues that constructive breakdowns can help experts to reflect and to externalize practical knowledge \cite{Schon1983}.
{I} have used this effect in 
{my} observations, e.g. when an unexpected question from a student or examiner ''interrupted'' supervision work and allowed 
{me} to augment any guidelines with previously tacit knowledge.
It is 
{my} hope to facilitate critical discussions by providing a language, shared sub-objectives, and knowledge of common pitfalls.

A major quality criteria for quality learning is the constructive alignment \cite{Biggs2007}, which demands that assessment in pedagogics strongly relates to learning objectives.
As 
argued in the introduction, the assessment of master's theses focuses on academic contribution, thus leaving it entirely to the student and supervisor to balance assessment on academic quality with practical relevance of the work done during the thesis. 
Especially in applied fields (such as software engineering and requirements engineering), academic excellence can be related to applying strong empirical methods \cite{Ralph2021}. 
In order to get access to empirical data from practice, companies must be convinced that they will get value in return. 

Given the objective (write a master's thesis that includes a scientific contribution) and the assessment (academic quality of research and writing), the critical question is whether students are provided with sufficiently strong means to actively pursue thesis work with a high chance for success. 
A high number of successful theses show that it is not impossible; good results have been achieved with qualitative exploratory studies \cite{Yin2014,Creswell2017,Runeson2009}, that investigate a new problem-space, or quantitative research to investigate causalities in practice \cite{Wohlin2012}.
It is however a big challenge for students and supervisors, to setup a thesis that aims to solve a practical problem.
Success depends on having sufficient knowledge on the team.

\input{hevner-guidelines-tbl.tex}

When considering knowledge in pedagogics, Koehler has suggested to distinguish technical, pedagogical, and content knowledge \cite{Koehler2008}.
In 
{this study's} particular context, \emph{technical knowledge} relates to technologies, methods, and practices for conducting master's thesis work. 
{The} basics {are generally} well-covered, since 
students have training in technical writing, academic integrity, and general research methods.
What is lacking though is knowledge about interacting with industry partners, about constructive research methods, and pragmatic guidelines for applying them.
\emph{Pedagogical knowledge} manifests in a set of workshops that support master's thesis works, in feedback given on proposals, halftime presentation, and final presentation as well as individual supervision. 
Multiple teachers are involved, yet mechanisms for aligning their views are only weakly developed. 
{It is the ambition of this paper to } contribute to improved alignment about constructive thesis works. 
\emph{Content knowledge} relates to the concrete topic of a thesis and while a supervisor or examiner should have sufficient background in the topic, the thesis itself should advance and increase the content knowledge. 
{This paper provides} specific knowledge about how reoccurring problems can be solved.

{The work presented in this paper} rests on a huge body of related work on empirical research methods.
Design science is very successfully applied in many disciplines, particularly in information systems research \cite{vaishnavi,wieringa2009,hevner}.
Increasingly, 
applications 
{are also found} in software and requirements engineering, yet there is a lack of examples as well as of topic-related tailoring of the methods.
A particular difference in the design science research 
{across those research fields} lies in the type of artifact: While established method descriptions aim to produce information systems that support human-machine interaction to solve a particular problem \cite{hevner}, 
{the aim within software and requirements engineering is often} to design methods and ways-of-working to solve 
{reoccurring} problems in 
{a specific} context. 

Thus, 
{this} work aims to complement existing works \cite{vaishnavi,wieringa2009,hevner,Storey2017,Engstrom2020}
{. S}tudents as well as supervisors and examiners are strongly recommended to read 
{these fundamental works before relying on the guidelines in this paper}.
In particular, 
design science research guidelines suggested by Hevner et al. 
are fundamental for any design science thesis \cite{hevner}.
Table \ref{tab:hevner} relates their guidelines to 
context{ in this paper}.

{The} guidelines {in this paper} are strongly influenced by Wieringa et al.'s suggestion of a \emph{regulative cycle} \cite{wieringa2009}, where researchers iteratively investigate the problem, suggest and validate solution proposals, implement a solution, and evaluate it. 
This is for example visible in  Guideline G2 to work in iterations as well as in Guideline G3, 
{which suggests} three questions related to the problem, the potential solutions, and their evaluation.
For exactly these areas, Storey et al.'s graphical abstract for design science research 
{demands for} rigorous methods \cite{Storey2017}.


To conclude, 
{this paper systematically presents} experience on running master's theses based on design science research as a complement to existing guidelines. 
{Its key contribution is to} provide value in the critical pedagogical discourse about running and assessing constructive master's theses.

%% file: hevner-guidelines-tbl.tex
\begin{table*}[th]
\caption{Hevner et al.'s guidelines for design science research and their relation to our context.}
\label{tab:hevner}
\centering
\begin{tabular}{v{0.45\textwidth}v{0.45\textwidth}}
\toprule
\textbf{Guideline by Hevner et al.} & \textbf{Applicability in this context} \tabularnewline
\midrule
\emph{Guideline 1: Problem relevance. }Hevner et al. argue that design science research can and should be applied \emph{``to develop technology-based solutions to important and relevant business problems''} \cite{hevner}. 
& I agree widely with this guideline; especially with the need of a relevant problem. 
However, I do not follow the constraint that a solution must be technology-based. 
I found design science research well applicable to the design of frameworks and strategies. 
\tabularnewline
\grayrow \emph{Guideline 2: Research rigor.}  Hevner et al. argue that \emph{``design science research requires the application of rigorous methods, both in construction and evaluation of the design artifact''} \cite{hevner}. 
&
I agree, but cannot fail to raise the issue that few methods exist for constructing a design artifact, beyond rough descriptions of agile or iterative approaches. 
Thus, I focus more on the interaction of students with supervisors from academia and industry during construction as well as on iterative work on the design artifact, so that evaluation results can positively affect construction. 
\tabularnewline
\emph{Guideline 3: Design science research as a Search Process.}  Hevner et al. describe design science research as heuristic search strategy, often based on decomposition of the initial problem \cite{hevner}. 
They argue that it is rarely possible to find the optimal solution to a relevant problem and that design science research instead should focus on finding any effective solution instead.
& This guideline can be confusing to students. I usually interpret it as a motivation to work incrementally and iteratively. 
\tabularnewline
\grayrow \emph{Guideline 4: Design as an Artifact.} I agree with Hevner that the effective representation of an artifact is key to its evaluation, since it allows implementation and application in its intended context  \cite{hevner}. 
& 
Our usual interpretation is that emphasis should be given to understanding the problem and to clearly relate any effort in construction or evaluation to the current understanding of the problem (e.g. through formulation of research questions). In addition, I recommend a dedicated section for elaborating on the artifact.
\tabularnewline
\emph{Guideline 5: Design evaluation.} 
Evaluation of the solution being fit-for-purpose and actually able to address the problem is key.
& 
The design evaluation methods listed by Hevner et al. are very useful \cite{hevner} and I usually push for observational evaluation. 
Since I do not necessarily follow Hevner et al.'s emphasis on technology-based solutions (their Guideline 1), the criteria of \emph{``aesthetically pleasing''} designs \cite{hevner} is often hard to operationalize. 
\tabularnewline
\grayrow \emph{Guideline 6: Research contributions.} 
Hevner et al. suggest that good design science research contributes to \emph{``the areas of the design artifact, design construction knowledge, and/or design evaluation knowledge''} \cite{hevner}. 
& 
Examiners and reviewers in our field often do not appreciate the knowledge about the concrete artifact as a stand-alone contribution, especially if it is a software tool.
I argue that in many cases, a crisp description of the problem can be a scientific contribution, thus I emphasize its value by suggesting a dedicated research question.
\tabularnewline
\bottomrule
\end{tabular}
\end{table*}

%% file: 03-method.tex

\section{Research method}
\label{sec:research-method}

The research method of this paper is inspired by design science \cite{vaishnavi,wieringa2009,hevner,Engstrom2020}.
In contrast to action research, design science does not focus on a particular action and its effects when introducing it to a case study, but instead emphasizes the actual design process, the creation of a design artifact \cite{Pfeffers2007}. 
While constructing the design artifact, the method ensures an empirical approach to design choices, which in turn allows answering certain knowledge questions.
 
The research questions in this paper are a good fit for the design science method. 
The artefact consists of guidelines for students and supervisors, in the beginning mainly driven by the personal need of the author and his students. 
This allows to better characterize the actual problem (RQ1), which was not as clearly visible from the beginning. 
By trying out different approaches and discussing others, 
potential mitigation strategies (RQ2) {are uncovered} for those problems.
{Finally, the }experience with useful mitigation strategies (good practices) and common pitfalls (RQ3){ is reported}.

Inspired by the regulative cycle \cite{wieringa2009}, the artefact has iteratively evolved, allowing  to refine 
{the} knowledge with respect to each research question.
In each cycle (or: iteration), 
{there is an improvement of the} ability to articulate the problem to be solved, to identify and validate solution candidates, to implement improvements of the guidelines for the next generation of thesis projects, and to evaluate the situation after concluding the theses in one cycle, before starting with the problem identification of the next cycle.

\subsection{Data collection}

The results in this paper are based on personal experience from several student theses. 
Table \ref{tab:sources} gives an overview of master's theses as well as papers derived from them. 
{T}heses {are grouped} by years and 
each year {is referred to }as a cycle. 
Thus, experience from previous cycles is available for those that follow.
For each master's thesis, data sources include the author's personal notes, discussions with students and colleagues, the actual theses and publications, 
and feedback from examiners and reviewers. 
Especially the personal notes and discussion are of informal and subjective nature, yet by triangulating them with the published theses and publications, 
{they become reliable enough and} valuable in the overall process of constructing useful guidelines. 

In addition, in the 2020 cycle, 
{the} guidelines {were presented} within a Research Method class in the Software Engineering and Management Bachelor Program at University of Gothenburg, discussed 
{with} interested faculty and PhD students, and 
{checked for} usefulness as well as understandability 
{with a survey}.
The survey was available as an online survey and distributed to participants in the research method class as well as to faculty and phd students.
Master's theses in 2020  had access to a draft of this paper and were also asked to provide feedback both orally and as part of the survey. 
In total, 10 surveys were filled out, of which 
{one was} invalid. 
Among the remaining participants were two Master students, two PhD students, four on a postdoctoral level (incl. faculty), and one who ticked ''other''.

\input{tab-sources}

\subsection{Data analysis}

For data analysis, 
information from different data sources {was grouped} along similar concerns with relevance to students and supervisor.
During supervision in each cycle, 
{I} looked out for similarities in these concerns and provided examples from previous cycles as part of supervision meetings. 
This helped 
to refine the advice and to see patterns of typical problems in such thesis projects, to list promising practices, and to learn about which {guidelines} actually work in favour of thesis projects.
{The results are reported} in a detailed narrative style to make this iterative data analysis as transparent as possible.
Visualizations and findings from the survey are provided to further illustrate 
{the} findings (mainly relevant for Research Question 3).

%% file: tab-sources.tex
\begin{table}
\caption{Overview of data sources (JP - journal paper, CP - conf. paper, MT - Master's thesis, BT - Bachelor thesis)
}
\label{tab:sources}
\begin{tabular}{lv{0.12\columnwidth}v{0.45\columnwidth}v{0.15\columnwidth}}
\toprule
\emph{Cycle} & \emph{Data type} & \emph{Artifact} & \emph{Source}\tabularnewline
\midrule
2014 & MT, CP & Feedback system (integrated in test car) to facilitate requirements feedback flows in iterative development & \cite{AR2014,KARI2015}\tabularnewline
\midrule
2015 & MT &  Concept and IDE plugin to improve unit testing practices with gamification & \cite{AJ2015}\tabularnewline
2015 & MT, CP & Automation of validating non-functional test results & \cite{IF2015,FIL+2016}\tabularnewline
\midrule
2016 & MT & Testing strategies to support continuous integration for complex systems & \cite{GJ2016} \tabularnewline
2016 & MT, JP & Loco Coco: Low Cost construction of Coordination and Communication networks from model-based systems engineering data & \cite{Moh2016,Mohamad2017}\tabularnewline
\midrule
2017 & MT& Test automation strategy to enable continuous integration for an automotive platform & \cite{SB2017}\tabularnewline
\midrule
2018 & MT & Requirements elicitation approach for intelligent and interactive systems in autonomous vehicles & \cite{SC2018}\tabularnewline
\midrule
2019 & MT & Architecture Framework for Blockchain Implementation & \cite{LS2019}\tabularnewline
2019 & CP, MT, BT & Concept and tool-support for managing textual (system) requirements in git (T-Reqs) & \cite{Knauss2018,Gebremichael2019,AB2019} \tabularnewline
\midrule
2020 & MT & Requirements engineering approach for developing cloud based support of autonomous trucks & \cite{Supriya2020}\tabularnewline
2020 & MT & Flexible integration concept of voice recognition components in an automotive android platform & \cite{JE2020} \tabularnewline
\bottomrule
\end{tabular}
\end{table}

%% file: 04-guidelines.tex
\section{Guidelines}
\label{sec:guidelines}



This paper is written as a design science research paper with the \emph{guidelines for running a design science research master theses with industry} as the artifact.
{This section presents} the seven guidelines (G1-G7), i.e. the artifact, concisely. 
Section \ref{sec:results} elaborates how these guidelines follow from our research method and empirical data.
While this order may appear backward, 
it often can be more elegant, since  {it allows to write }the empirical results 
with the final version of the artifact in mind
{. It also allows} the reader 
{to} refer to the artifact section for guidance (see G7).




\begin{summarybox}
(G1) Define the \textbf{artifact} early. 
Use it to agree with all stakeholders on the thesis goals with respect to knowledge contribution.
\end{summarybox}

Design science helps to answer knowledge questions through creating a concrete design artifact. 
Often, the exact nature of the design artifact is hidden in the beginning. 
Still, students and industry partners need to agree what should be constructed during the thesis. 
The artifact needs to guide the knowledge questions, not the other way.
The artifact can evolve over time, but it is important to always have the best knowledge about the artifact documented explicitly.

\begin{summarybox}
(G2) Work in \textbf{iterations}. Improve the artifact and the knowledge in each iteration. Specifically, contribute to each research question in each iteration. 
\end{summarybox}
Each design science cycle corresponds to one iteration in which the artifact should be improved. 
A typical Master thesis can achieve three full cycles, each of them lasting a month. 
For achieving rigor, it is important that new knowledge is generated for each research question (by executing each phase of the regulative cycle \cite{hevner}) in each iteration.
Iterations help to manage incomplete knowledge about the exact problem and about how a suitable solution could look like. 
Some students try to answer one research question per iteration, but 
{this approach fails to leverage the benefits of iterative work towards the knowledge questions}. 

\begin{summarybox}
(G3) Define \textbf{research questions} with respect to the regulative cycle, i.e. one related to the problem, one related to potential solutions and their construction, and one related to evaluation (match solution to problem).
\end{summarybox}
Each design science cycle corresponds to one iteration in which the artifact should be improved. 
A good starting point is to define three research questions that correspond to the phases of the regulative cycle \cite{hevner}.
Consider an artifact that should support a company in doing X.
Then good initial questions are: 
\begin{description}
\item [RQ1] What is the problem with the current way of doing X?
\item [RQ2] Which potential solutions could mitigate this problem?
\item [RQ2.1] (optional) Which potential solutions are most promising with respect to potential value / effort of implementing?
\item [RQ2.2] (optional) How can a suitable solution be implemented in the artifact?
\item [RQ3] To what extent can the problem with the current way of doing X be solved by potential solutions?
\end{description}

The main questions 1-3 are all critical. 
RQ1 provides enough detail so that RQ3 can evaluate the quality in use, thus support the need to answer empirical knowledge questions. 
RQ2 (and its subquestions) aims to support the refinement of the artifact. 
Consider an Ishakawa (or fishbone) diagram to describe the problem and maintain a strong synergies between answering research questions and developing the artifact.

\begin{summarybox}
(G4) Have regular \textbf{meetings}. Student and supervisor meet once per phase (once per week). Student, industry supervisor, and academic supervisor meet once per cycle (once per month). Goal: support rigor.
\end{summarybox}
Students and supervisor should meet at least once per phase of the regulative cycle, which resonates well with a typical weekly or biweekly supervision rhythm.
In addition, students, industry representatives, and supervisor should meet before each cycle to plan activities.
Without regular meetings, problems and misunderstandings can go unnoticed for too long.
Often, it is very difficult for students to moderate these cycle-meetings, since they lack experience in design science research. 
Especially in the first meetings, 
the supervisor 
{should} take an active role to align and balance the different perspectives of students, industry, and academic examination.

\begin{summarybox}
(G5) \textbf{Shift emphasis} between cycles. Work on each research question in each cycle, but put more emphasis on RQ1 (Problem) in Cycle one, on RQ2 (Solutions) in Cycle two, and on RQ3 (Evaluation) in Cycle three.
\end{summarybox}
The first cycle requires much more effort with respect to understanding the exact problem to solve. 
Still, make sure that already now solutions are provided and evaluated.
Focus on implementation in the second cycle, but be prepared to update the understanding of the problem and to test your implementation through evaluation. 
Use the third cycle for polishing, but as the last cycle, it should focus on the evaluation. 

It is important to describe very clearly which data was collected when. 
Table \ref{tab:data-per-cycle} provides a proven structure for this purpose.
Since the RQ typically should align with the phases of the regulative cycle, data collection is presented in relation to RQs.

\begin{table}
\centering
\caption{Template for a table that reports which data was collected in which cycle (G2) with respect to each RQ (or corresponding phase of regulative cycle (G3). The gray area shows the shifting emphasis between cycles (G5) and the continuous work on the artifact (G1). Typical examples of data collections are provided.}
\label{tab:data-per-cycle}
\includegraphics[width=\columnwidth]{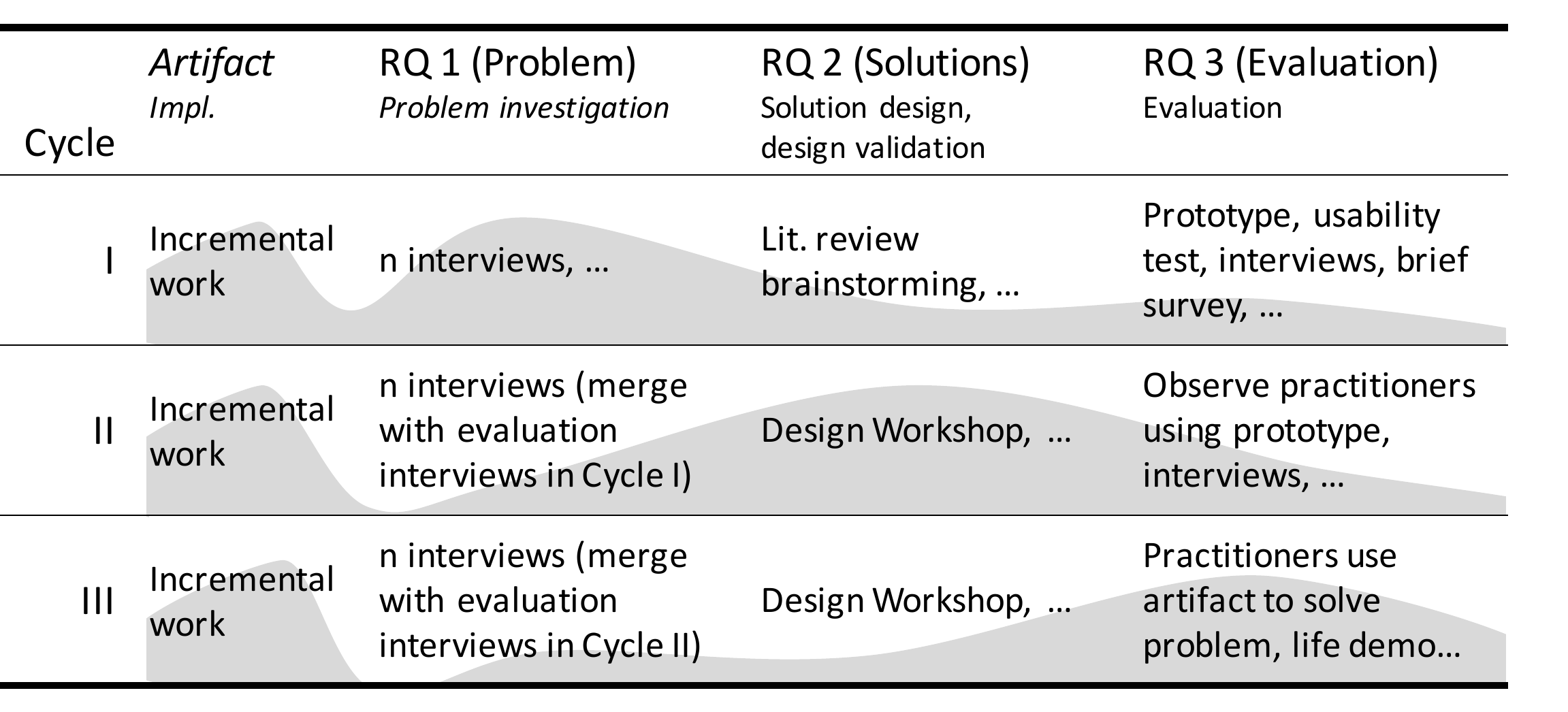}
\end{table}

\begin{summarybox}
(G6) Have a dedicated section to \textbf{describe the artifact}. A concise description of the artifact is beneficial to the casual reader and allows other sections (e.g. findings, discussion) to focus on the learning, not the artifact. 
\end{summarybox}
When coming from an empirical background, it is tempting to press all content either in a findings section (usually any objective observations derived from data) and in a discussion section (to more subjectively discuss the implications of the findings).
{W}ith design science thesis work, this results in very complicated descriptions in both sections. 

It is much better to concisely describe the final artifact, as it was after the final cycle. 
It is very helpful to give the artifact a clear name, so that it can be referenced throughout the paper. 
It can be useful to present the artifact early on, even before the findings. 
The findings are then only presenting support for why the artifact is constructed as it is, allowing to clearly relate any observations to elements in the artifact. 
This is however not always the case. 

\begin{summarybox}
(G7) \textbf{Write the thesis} document as you go, but do not submit it that way. 
Often, the best way to report your results can only be decided after Cycle 3. 
Thus, chose the least obstructive way of documenting the results during the thesis (chronological). Then restructure before submission.
\end{summarybox}

Since the work is done iteratively, the structure of the thesis is complicated. 
Do we report problems, solutions, and evaluation separately per cycle (or iteration)? 
Then, it is hard to derive the overall answer for the main research questions from the text. 
Do we structure the findings primarily by research questions? 
That might be hard to do during the iterations, as it is easier to keep the observations from one cycle close together. 
Finding a suitable abstraction is much easier to do after the last cycle concludes. 
Also, the specific activities in Cycle II and III might depend on the findings 
{from} 
a previous cycle and it can be hard to create a consistent report that does not follow the chronological order of events.

Thus, 
students {should} write a thesis document as they go, where they fill in the findings for each Cycle in chronological order {(see left column in Table \ref{tab:structure})}.

This is, however, most often not a suitable structure for future readers of the thesis or paper. 
Thus, 
some time {should be scheduled} to restructure between the last cycle and submission of the paper (see right column in Table \ref{tab:structure} for a possible structure).
Students should rely on their supervisor's experience for identifying the best final structure, which can be a mix (as for example chosen in this paper).

\begin{table}
\centering
\caption{How to structure the report.
}
\label{tab:structure}
\input{tab-structure}
\end{table}

%% file: tab-structure.tex
\begin{tabular}{v{0.4\columnwidth}v{0.4\columnwidth}}
\toprule
\emph{While working in Cycles} & \emph{When submitting to examiner} \tabularnewline
\midrule
1. Introduction
& 
1. Introduction
\tabularnewline

\grayrow 2. Related work
& 
2. Related work
\tabularnewline

\rowcolor{white}
3. Research method \\
~3.1 Problem\\
~3.2 Solution candidates\\
~3.3 Validation\\
~3.4 Implementation\\
~3.5 Evaluation
& 3. Research method \\
~3.1 Problem\\
~3.2 Solution candidates\\
~3.3 Validation\\
~3.4 Implementation\\
~3.5 Evaluation
\tabularnewline
\midrule
\graycell 4 Findings\\
~4.1 Cycle I Findings\\
~~4.1.1 Problem\\
~~4.1.2 Solution candidates\\
~~4.1.3 Validation\\
~~4.1.4 Implementation\\
~~4.1.5 Evaluation\\
~4.2 Cycle II Findings\\
~~4.2.1 Problem\\
~~4.2.2 Solution candidates\\
~~4.2.3 Validation\\
~~4.2.4 Implementation\\
~~4.2.5 Evaluation\\
~4.3 Cycle III Findings\\
~~4.3.1 Problem\\
~~4.3.2 Solution candidates\\
~~4.3.3 Validation\\
~~4.3.4 Implementation\\
~~4.3.5 Evaluation
& 
4. The Artifact
\tabularnewline
\midrule
5. The Artifact
& 
\graycell 5. Findings\\
~5.1 RQ1 (Problem) \\
~5.2 RQ2 (Solution candidate) \\
~5.3 RQ3 (Evaluation) \\
\tabularnewline
\midrule
\grayrow
6. Discussion\\
~6.1 Answer to RQ 1 (Problem)\\
~6.2 Answer to RQ 2 (Solution)\\
~6.3 Answer to RQ 3 (Evaluation)
& 
6. Discussion\\
6.1 Implications for research\\
6.2 Implications for practice
\tabularnewline
\midrule
7. Conclusion and Outlook
&
7. Conclusion and Outlook
\tabularnewline
\bottomrule
\end{tabular}

%% file: 05-results.tex
\section{Results}
\label{sec:results}

This section presents findings with respect to the research questions based on seven years of experience with running design science master's theses. 
The findings relate to typical problems (RQ1), potential solutions (RQ2), and any evidence that such solutions actually solve the problems (RQ3). 
The knowledge for each research questions has evolved over the years and this section presents this evolution as a narrative supported by concrete data and experiences.

\subsection{Findings per cycle}


{This section reports on how DSR experiences and guidelines have evolved over time. }

\subsubsection{Findings from 2014 Cycle}
In the first thesis example \cite{AR2014} during the 2014 cycle, the company partner wanted the students to implement a way that allows to use feedback from test-drivers in developing SW-intense vehicle functions more systematically. 
In fact, a detailed analysis of the given problem (enable systematic feedback from test-drivers) from different stakeholder perspectives (developers, test-drivers) yielded surprises, even for the industry partner.
For example, tracing a given issue or test-driver report to a specific test-vehicle and to determine which software version was installed on that vehicle during the time of the test-drive was found to be very time-consuming and error-prone.
Also, the memory of which events occurred during a test-drive and in which situations proved to be surprisingly inaccurate. 
Our \emph{expectations on which topics this master's thesis would touch changed significantly} during the first iteration.
From this experience, we derive the recommendation to work in \textbf{iterations (G2)}.

One problem that we started to recognize already in this first thesis example is on \emph{how to report on iterative work and how to structure the thesis}. 
During the work, it is most appropriate to report by cycle and phase. 
For the reader, this can be very repetitive and the overall findings are less than clear. 
Also, it takes a lot of space, which ultimately let the reviewers of a conference publication of this master's thesis to suggest massive shortening and to accept the resulting paper only as a short paper \cite{KARI2015}, in which we mention the iterative approach only very briefly.
This experience contributed to 
Guideline \textbf{(G7)} on structuring the report.  

\subsubsection{Findings from 2015 Cycle}
In the 2015 cycle, {I supervised} two theses 
that employed a design science method.
While the artifact for one thesis was very clear from the start (since previous work had already established gamification concepts for unit testing), it was not as clear for the second thesis. 
In fact, only our ambition to characterize the artifact early based on frequent meetings allowed us to avoid larger problems in this thesis. 
In hindsight, even though the thesis and overall research was very successful, we should have been stricter with 
{scheduling joint meetings with} industry participants
{,} students{, and academic supervisor} early on. 
Instead, a series of bilateral meetings was scheduled with increasing frequency, when challenges for the thesis became clear. 
It was only in the second regulative cycle, that we stopped to think about this work as design science research - understanding the actual problem took so much time, that we instead reported the thesis as an exploratory qualitative case study.
This detailed problem description was found so valuable that we were able to bring the thesis to a very successful end and to publish it at a very good conference on testing.

From this cycle, 
{I} derive the need to have regular meetings (Guideline \textbf{(G4)}), to define the artifact early (Guideline \textbf{(G1)}), and to define research questions with respect to the regulative cycle (Guideline \textbf{(G3)}), especially since we came to value a clear description of the problem. 
The contrast in how well the two theses worked out 
{convinced me } that a clear artifact (Guideline \textbf{(G6)}) relating to a clear problem creates a very solid starting point. 
If both are given, the need for regular meetings with industry might be less pronounced.

\subsubsection{Findings from 2016 Cycle}

In 2016, 
{I supervised} two design science research theses, leading to one publication in a top journal. 
In 
{both} theses, the guidelines 
{identified} so far were very valuable. 
The thesis we struggled most with again emphasized problems when the artifact is not clearly described early on. 
It however also showed that a very abstract artifact can be created, here a testing strategy for continuous integration \cite{GJ2016}. 
Also, in 
{both} theses of this cycle, we struggled to define research questions and to manage different themes for each sprint. 
Thus, 
{I started to consider } recommendations for guidelines concerning the research questions (Guideline \textbf{(G3)}, especially: do not change them between cycles!)
 and to shift the emphasis between cycles (Guideline \textbf{(G5)}). 
 
 In one thesis \cite{Moh2016}, we aimed for generalization beyond the company context in the last iteration. 
 While it worked fine here, it also resulted in much additional work and a risk to generate less valuable data. 
 Therefore, 
 {I} have not raised this practice into 
 {the} guidelines.
 
 \subsubsection{Findings from 2017 Cycle}
 
 Encouraged from our success in 2016, 
 {I} dared to apply design science research in a more complex scenario, a testing strategy for a whole platform. 
 One of the learnings here was that in a complex situation, it can be very challenging to keep a good overview and to present the status in a consistent way. 
 At halftime of the thesis, the examiner was unsatisfied with the level of detail with respect to academic background and the repetitive, chronological state of the report. 
 While we did not disagree with his assessment, we stuck to our priorities and focused on getting the artifact under control as a highest priority \textbf{(G1)} and to fix the academic writing later \textbf{(G7)}.
 This was risky, since a lack of strong background (here on theory of testing) might produce an invalid artifact. 
 We took this risk, trusting that the artifact was not in conflict with established testing theory. 
 With this assumption in mind, we found it more likely that the quality of the academic writing can be improved later on, while we would not be able to retrospectively redefine the artifact. 
 
 Despite the challenges, students, examiner, and supervisors at university and company agreed that in the end a successful thesis was produced. 
 {I} take it as confirmation that 
 {the guidelines become more reliable and} increase 
 {the} ability to manage design science research theses even in complex situations. 
 However, 
 {I} will aim to keep examiners that are not experienced with 
 {my} approach better informed and 
 {I} realize that a thesis in a challenging environment will continue to be challenging, despite 
 {the} guidelines.

  \subsubsection{Findings from 2018 Cycle}
 
 The 2018 thesis worked out very smoothly, and we were able to apply all of our guidelines. 
 Again, the thesis addressed a very complex topic, but since we could rely on the case company's research department for industry supervision, it was easier to build a bridge between academic and pragmatic views. 
 A publication of this master's thesis is still forthcoming. 
 In compiling that publication, a concise description of the artifact was again a  main challenge, even though the students had focused on it in good time. 
 
 \subsubsection{Findings from 2019 Cycle}
 
 Our experience in 2019 can again be seen as a confirmation of our guidelines, which were applied and found useful. 
 However, with respect to T-Reqs, we struggled to embed the thesis in prior and parallel work. 
 {I} list these challenges below as known open problems, on which 
 {I} need more experience to provide clear guidance to supervisors and students. 

\subsubsection{Findings from 2020 Cycle}
In 2020 we approached two tough problems in the automotive sector, where industry partners aimed to gain new insights in challenging areas.
Consequently, the problem definition was open-ended and somewhat vague at the beginning.
In addition, one industry partner seemed slightly more interested in identifying suitable future employees than in solving a concrete problem. 
As a result, we struggled to concretise requirements for the artifact, as initial interviews were not sufficiently constrained and revealed a broad set of potentially relevant challenges.

We decided to use Ishakawa diagrams to mitigate these challenges. 
The diagram allows to relate challenges to the thesis topic in a tree-like fashion and to pick a specific challenge or subtree when designing or evaluating a solution. 

Since not all challenges in the diagram are addressed in the thesis, readers and examiners found it slightly unfocused. 
We attempted two different strategies to address this challenge: (i) make explicit which parts are out of scope and why; (ii) redraft the diagram to only show the area of focus.

While more experience is needed before suggesting a concrete guideline, 
{I} recommend to consider a formal, rigorous (qualitative) model when reporting on the problem in relation to \textbf{(G3)}.
 
 \begin{figure*}
\includegraphics[width=\columnwidth]{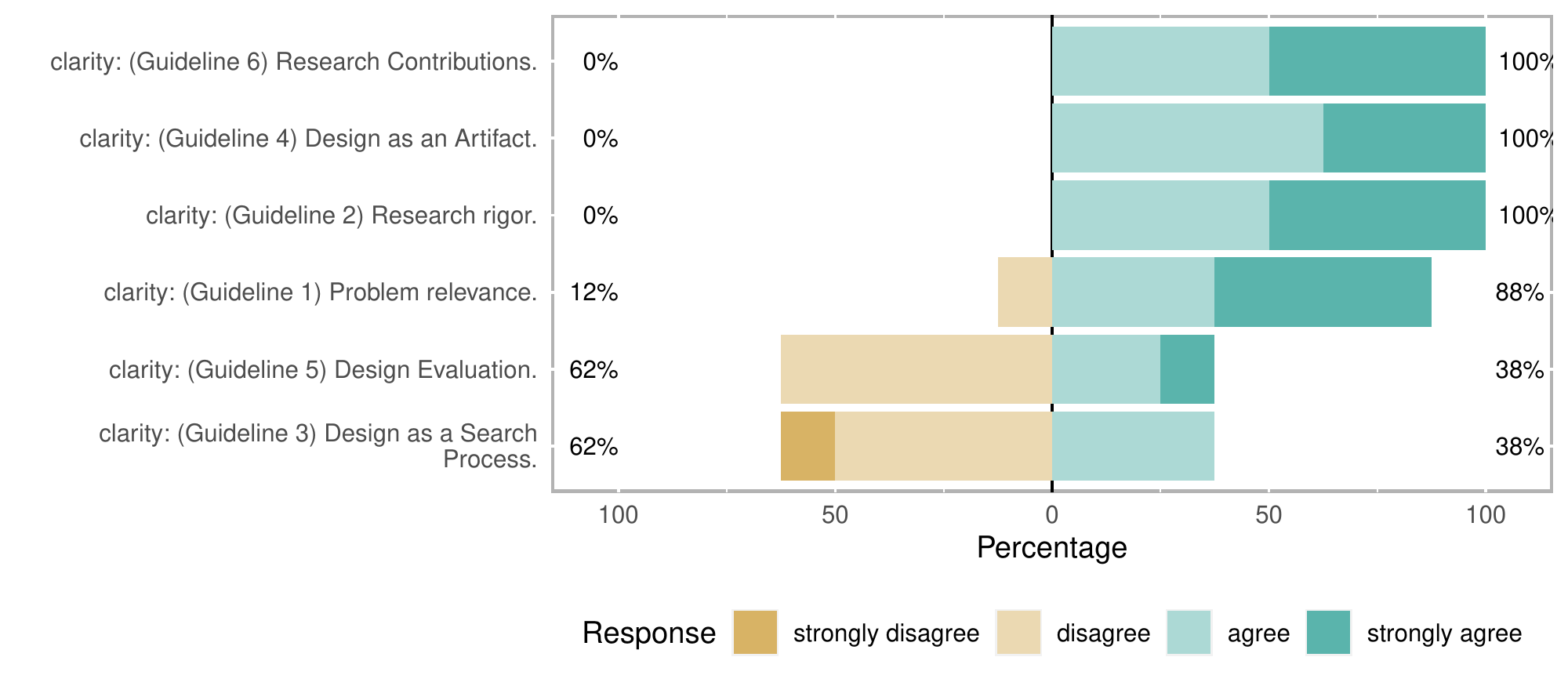}
\includegraphics[width=\columnwidth]{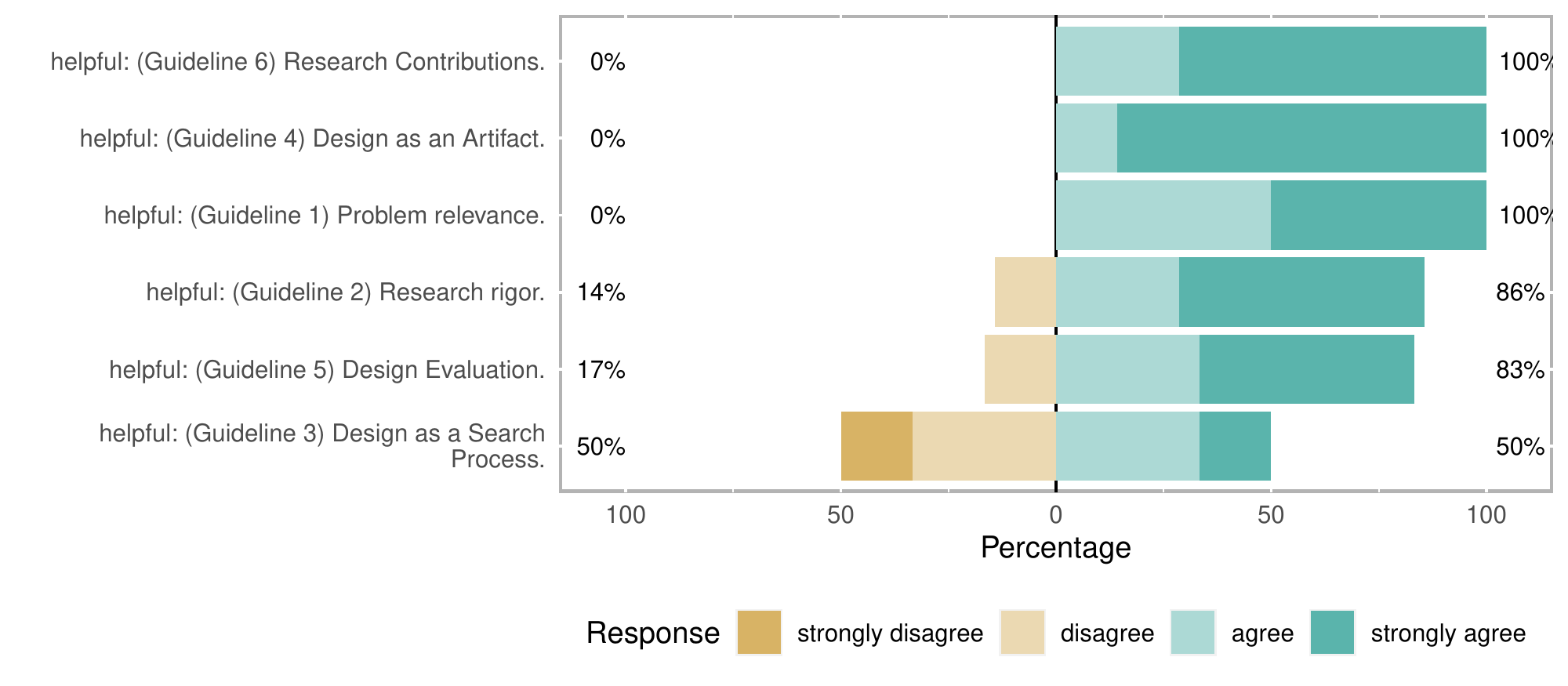}\\
\includegraphics[width=\columnwidth]{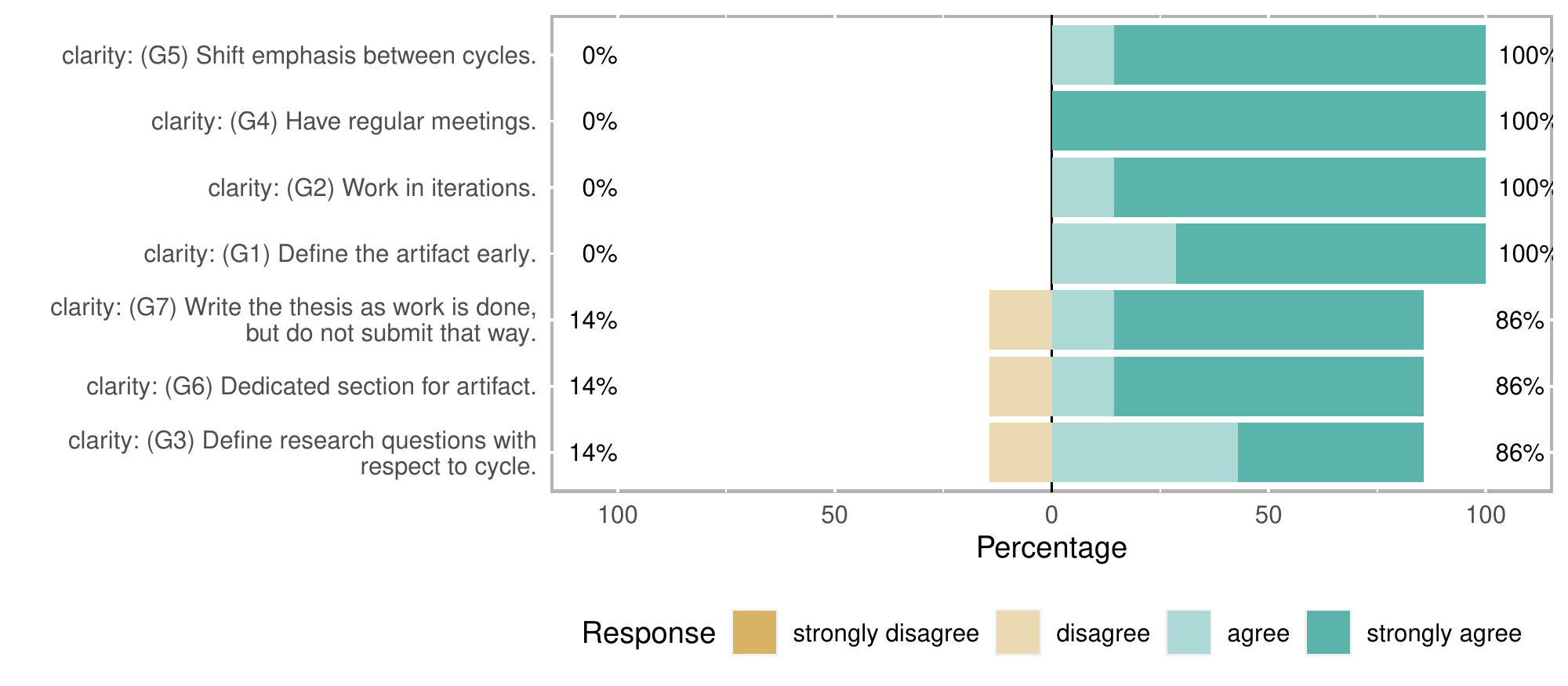}
\includegraphics[width=\columnwidth]{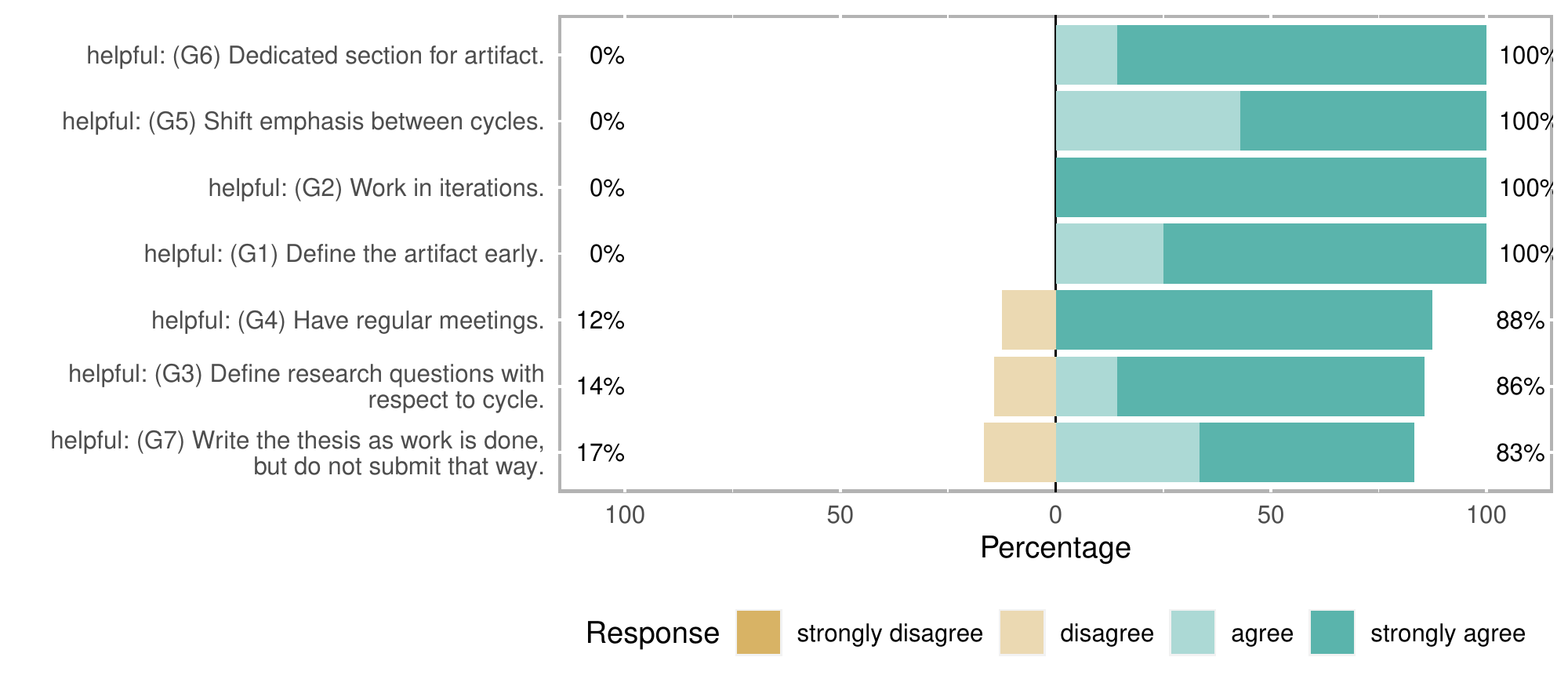}
\caption{Answers to likert scale questions: Do you agree that Guideline x from Hevner et al. \cite{hevner} (top row)/ from this study (bottom row) is clear (left column) / helpful (right column). The original likert statements included a short representation of the guideline in line with the representation in this paper.}
\label{fig:survey}
\end{figure*}

\subsection{RQ1 -- What problems exist with running a design science master's thesis?}

The {design science research method} 
is a potential solution to a re-occurring problem in any applied research domain that facilitates field work during master's thesis:
\begin{itemize}
\item How can a master's thesis do constructive work with relevance to practitioners and still be evaluated positively in an academic setting that favours empirical work?
\end{itemize}

This problem is related to a common misconception among students and industry, that has been formulated in different variants of  \emph{``Industry wants master's students to do programming tasks, academy does not allow programming''}. 

Other {reoccurring} problems 
are the following:

\begin{itemize}
\item Practical problems are often not well understood at the start of the thesis. This relates to all stakeholders: The students as well as academic and industry supervisor.
\item Working in a complex industry environment on a practical problem that is not well understood is a hard problem. 
\item It is difficult to report on iterative work and to find a good structure for the thesis.
\item It is hard to fully leverage the different perspectives of all stakeholders to properly manage risks in thesis work. 
\item It is hard to plan iterative research cycles and to frame good research questions. If research questions relate to only a specific iteration, there is a danger to not get enough data or good enough quality of data. 
\end{itemize}

\subsection{RQ2 -- Which potential solutions exist to mitigate typical problems with running a design science master's thesis?}

As demonstrated, 
{the} guidelines {in this paper }are derived from the experiences over recent years. 
Beyond the guidelines reported in Section \ref{sec:guidelines}, further potential solutions were collected, but not (yet) added to the guidelines, since 
{I am} not convinced that they 
{can} be generalized:
\begin{itemize}
\item Use one cycle to generalize the artifact beyond the case company. 
{I} think that there is a high risk to gain little data of sufficient quality, which must be carefully considered with respect to the potential gain.
\item Include examiner in discussions on research plan and trade-offs.
While communication is important, we rely on separating supervisor and examiner role in our context.
It is clear that ambitious thesis projects will bring both roles out of their comfort zone and we did not experience problems that would demand immediate action, i.e. getting feedback of the examiner on proposal, half-time presentation, final thesis, and defense, is generally sufficient. 
\end{itemize}

Further, guidelines by Hevner et al. \cite{hevner} and the regulative cycle \cite{wieringa2009} should be considered as potential solutions.

\subsection{RQ3 -- To what extent can typical problems with running a design science master's thesis be addressed?}

In Figure \ref{fig:survey} we show the results from our survey. 
It confirms our impression that participants struggle with the ''original'' guidelines by Hevner et al. \cite{hevner}, especially with the important aspects of design evaluation and design as a search process. 
All guidelines, regardless of their source, were found helpful, and the results support our claim that our actionable guidelines complement the information available to software engineering students in a useful way.

This is further supported by quotes from survey and reflection with students from the 2020 cycle.
For example, one PhD student wrote \emph{''I do not get the activities I should perform from these guidelines [referring to Guidelines by Hevner et al.]. I however think they are helpful in checking that one is not below requird design science standards but not in doing the research itself.''}
In contrast, 
{the} guidelines {here} 
{target} students specifically and {were described} to be especially helpful to those who are not yet familiar with design science research. 
Some participants worried that guidelines might be too specific for our particular context, e.g. limiting the research questions as expressed in G3, while others (G4, G7) might be general advise and not specific to design science research.

Feedback from master's students in the 2020 cycle confirms that the guidelines help to get started.
In particular the step-wise presentation was seen as helpful.

It is 
{my} impression, that 
{the} guidelines have stabelized in recent years. 
They have been used in several successful theses and found 
{by myself and other supervisors} to mitigate the problems listed above {(see Table \ref{tab:problem-solution-mapping} for a mapping between problems and guidelines)}. 
They are not strictly necessary (different approaches can also create successful theses) and they are not complete (by following all guidelines, it is not guaranteed that the thesis is successful). 
As such, 
{I} believe that they are a valuable complement to other descriptions of design science research method, as discussed in Section \ref{sec:background}.


\begin{table*}
\caption{Mapping of identified problems (RQ1) to proposed guidelines (RQ2).}
\label{tab:problem-solution-mapping}
\begin{tabular}{v{0.4\textwidth}v{0.55\textwidth}}
\toprule
Problem & Applicable guidelines \tabularnewline
\midrule
How can a master's thesis do constructive work with relevance to practitioners and still be evaluated positively in an academic setting that favours empirical work? & The combination of all guidelines presented here helps to address this problem, specifically the iterative work on research questions (G3), the shifting of emphasis (G5), and the considerations when reporting results (G7) support solving this problem. \tabularnewline
\grayrow Practical problems are often not well understood at the start of the thesis. This relates to all stakeholders: The students as well as academic and industry supervisor. & It helps to define the artifact early (G1), to work in iterations (G2) with shifting emphasis (G5).\tabularnewline
Working in a complex industry environment on a practical problem that is not well understood is a hard problem. & The guidance provided in this paper can help to navigate this hard problem. In particular, aligning research questions with iterations (G3) and the attention given to understanding the problem, as well as the guidance in Table \ref{tab:data-per-cycle} can provide good starting points.\tabularnewline
\grayrow It is difficult to report on iterative work and to find a good structure for the thesis. & Guidelines G6 and G7 was introduced to address this problem. \tabularnewline
It is hard to fully leverage the different perspectives of all stakeholders to properly manage risks in thesis work. & Regular meetings aligned with DSR (G4) and the guidelines relating to artifact (G1, G6) and research questions (G3) support balancing these perspectives.\tabularnewline
\grayrow It is hard to plan iterative research cycles and to frame good research questions. If research questions relate to only a specific iteration, there is a danger to not get enough data or good enough quality of data. & Addressed by G3 and G7. \tabularnewline
\bottomrule
\end{tabular}
\end{table*}

%% file: 06-discussion.tex
\section{Discussion}
\label{sec:discussion}

\subsection{Limitations}

The aim of this study is to provide guidelines for master's theses in software engineering. 
Even more specifically, the author's experience relates to requirements engineering and development processes, topics that are usually less technical than other research in the broader software engineering domain. 
Results should reasonably well translate  to applied research, involving practitioners, in comparable domains. 
Yet, generalizability of results is not the main goal of this study.

Instead, this study aims to provide in depth knowledge and, in doing so, embraces also subjective interpretations of the author. 
This of course affects the reliability of results.
Special attention has been put on making any subjective evaluations of feedback explicit and to separate them from more objective observations. 
In addition, the process is described in detail, to support recoverability, that is, to allow other researchers to reason about why they might observe different results in a similar setting. 

Survey participation was low, especially among students. 
For them, it is hard to judge whether guidelines are clear or helpful before doing a thesis.
Participation of (potential) supervisors on PhD level and beyond was slightly higher, offering a good complement to other data collected in this study. 
Especially the free-text fields proved valuable feedback.

\subsection{Implications}
Thuan et al. suggest three types of research questions for design science studies \cite{Thuan2019}, relating to \emph{the way of knowing, the way of framing, and the way of designing}.
The research questions in our Guideline \textbf{(G3)} only partially match this view: Our suggested RQ1 (problem) relates to the \emph{way of framing}, and our suggested RQ2 (potential solutions) relates to accessing the knowledge base as well as to \emph{ways of designing}, which we do not distinguish in our guidelines. 
We suggest an additional question to explicitly evaluate the design in use.
In this way, we believe that our Guideline \textbf{(G3)} is in line with the generic graphical abstract that Storey et al. suggest for design science research in software engineering \cite{Storey2017}.
Our RQ1 (problem) relates to the problem understanding, RQ2 to the solution design, and RQ3 to the validation approach (or more precisely to the actual validation).

A key component of the graphical abstract proposed by Storey et al. is the technology rule, which itself is based on van Aken's work \cite{vanAken2005}: \emph{''To achieve $<$Effect$>$ in $<$Situation$>$ apply $<$Intervention$>$''} \cite{Storey2017}.
{The guidelines in this paper} are in line with this way of framing design science research explicitly, yet more experience in applying this template in our thesis work must still be gained.
In particular, it is unclear, how many technology rules can be suggested in one thesis project. 
One  {single rule} may be too rough for practical relevance (e.g. framing this study as ''To achieve practical and academic relevance in a software engineering master's thesis apply our Guidelines''), while several technology rules (e.g. one for each guideline in this paper) may be too unfocused for academic relevance.

Engstr{\"o}m et al. cluster design science research studies with respect to the graphical abstract into descriptive, solution-design, solution-validation, and problem-solution \cite{Engstrom2020}.
{The guidelines in this paper} aim for problem-solution studies that validate the solution's quality in use against a precise problem definition.
Comparing 
{the} guidelines with recommendations by Engstr{\"o}m et al. \cite{Engstrom2020} provides an additional validation of our guidelines. 

\begin{itemize}
\item \emph{Explicit design science constructs} are partially implied through our Guidelines 1-3 and 6. 
{I} recommend to use the graphical abstract in addition to 
{the} guidelines in future theses. 
\item \emph{Use real problem instances} is fulfilled through the context of our theses.
\item \emph{Choose validation methods and context} is implied by 
{the} guidelines, in particular by Guideline 3 and 5, which together allow to revisit the research question and design iteratively and to adjust the method to the context as it is revealed.
\item \emph{Use design science research lens as research guide} is 
{the} main goal in this paper. 
{I} claim that 
{this} study operationalizes this lens into concrete, actionable guidelines. 
\item \emph{Research design as design science} has been a central consideration for the design of this study. By applying the growing knowledge base of design science research method papers in a concrete context, 
{I} design 
{the} guidelines and hope that they, together with answers to 
{the} knowledge questions, in turn contribute to understanding and applicability of this method within software engineering. 
\end{itemize}
 
\subsubsection{Implications for students}

{My} general recommendation to students is to make use of the supervision team, including the academic and industry supervisor.
Use them as sounding board during evaluating solution candidates and when planning the different cycles. 

In addition, 
{I} recommend to provide value to practitioners early on, they are more likely to participate in your research,

\subsubsection{Implications for practitioners}

Clearly, a thesis becomes more valuable the more you invest into it. 
%
{The best practices here} indicate some ways on how to become active: 
\begin{itemize}
\item Help to clarify the actual problem, it is often much more complicated and hard to understand than you may believe. 
\item Provide access to different internal stakeholders and moderate the process.
\item Help the student to discuss the practical relevance of their work, and in particular make sure that the artifact is of value to you and your organization.
\end{itemize}

\subsubsection{Implications for supervisors}

{The} guidelines show the specific supervision needs of a design science research thesis in industry. 
The problem of balancing the trade-off between academically acceptable research method and practically relevant constructive work is too tough for the students or the industry partners to handle themselves.
It requires a hands-on supervisor, especially, since the constructive alignment of requirements towards the master's thesis, its learning outcomes, and its evaluation criteria often is not clearly visible for a constructive thesis.

\subsubsection{Implications for examiners}

It is hard to compare the amount of empirical data between design science research thesis and other empirical investigations. 
Usually, 
{I} found design science research interviews to be shorter and more structured. 
They can be less formal, since the artifact allows quick interaction with practitioners at random times.
Analysis is often more straight forward than in an exploratory qualitative case study, since it is constrained by the artifact. 

Still, by iteratively revisiting similar research questions, the quality of the data is very high and through the concrete artifact, 
{researchers (i.e. theses workers)} can provide deeper insights then with an interview-only study.
In addition, the effort for managing iterations (as indicated in these guidelines) and the effort for designing the artifact must be considered, when coming to a fair assessment. 

\subsection{Known open problems}

Finally, 
{I am} aware of certain challenges for which 
{I} have so far not found a good way of managing them.
I hope that future work will provide more input on these.

\subsubsection{What if the problem is not sufficiently solved after the thesis?}

Design science theses can be a lot of effort. 
It is possible that a thesis provides enough results to be acceptable, yet, for a scientific contribution or even to really solve the industry partner's problem, additional work is required. 
Yet, a thesis must provide a good scope.

We have tried to do additional iterations/cycles when moving from the thesis to a research paper.
Students have been hired by companies to see promising artifacts being fully implemented.
The problem is how a) to report the thesis as something complete, and at the same time b) to create awareness that additional work must be done. 
Depending on how these aspects are handled, it can be hard for a follow-up thesis to start the first cycle from the problem description of the previous thesis.
{I} do not have good advise for this situation at this moment.

\subsubsection{How to formalize problems that are initially too vague?}

When the initial scope is too broad, interviews might not have enough focus and it becomes hard for students to highlight the most relevant problems to be addressed by their design. 

When 
{I} ran into this challenge, 
{I} have tried to either {help students to} formalize the problem, e.g. through providing Ishakawa models, or to refocus the thesis on investigating the problem area rather than to design a solution.
Neither solution is without pain and future work will have to reveal elegant ways of solving this.


%% file: 07-conclusion.tex
\section{Conclusion and Outlook}

{This paper presents} actionable guidelines for design science research thesis work based on seven years of experience with conducting 12 theses.
{From this experience, seven guidelines were systematically derived and their evaluation indicates} that they complement existing research methods in the field and make design science research more approachable for students in the short time frame of a typical master's thesis. 
{There reamain t}wo known problems for which 
{guidelines still have to be established}, especially ways to continue work beyond an individual thesis and to formalize problem definitions for wicked or vague problems. 
{I} hope that 
{the work presented in this paper can facilitate} constructive work in collaboration with industry and by this strengthens the alignment of educational goals with realistic thesis projects. 